\documentclass[aps,prl,reprint,superscriptaddress]{revtex4-1}

\usepackage[T1]{fontenc}
\usepackage{amsmath}
\usepackage{graphicx}
\usepackage{xcolor}
\usepackage[noend]{algpseudocode}
\usepackage{algorithm}
\usepackage{qcircuit}
\usepackage{hyperref}
\usepackage[normalem]{ulem}

\usepackage{dsfont}

\makeatletter
\def\BState{\State\hskip-\ALG@thistlm}
\makeatother

\newcommand{\vect}[1]{\boldsymbol{\mathbf{#1}}}
\def\ket#1{\left| #1 \right\rangle}

\def\vev#1{\left\langle #1 \right\rangle}

\begin{document}

%\preprint{}

%\title{Scaling to larger quantum simulation problems using operator pool tiling in ADAPT-VQE}

\title{Scaling adaptive quantum simulation algorithms via operator pool tiling}


\author{John S. Van Dyke}
%\thanks{}
\altaffiliation[Present Address: ]{Johns Hopkins University Applied Physics Laboratory, 11100 Johns Hopkins Road, Laurel, Maryland 20723, USA}
\thanks{These two authors contributed equally}
\affiliation{Department of Physics, Virginia Tech, Blacksburg, VA 24061}

\author{Karunya Shirali}
\thanks{These two authors contributed equally}
\affiliation{Department of Physics, Virginia Tech, Blacksburg, VA 24061}

\author{George S. Barron}
\affiliation{Department of Physics, Virginia Tech, Blacksburg, VA 24061}

\author{Nicholas J. Mayhall}
\affiliation{Department of Chemistry, Virginia Tech, Blacksburg, VA 24061}

\author{Edwin Barnes}
\affiliation{Department of Physics, Virginia Tech, Blacksburg, VA 24061}

\author{Sophia E. Economou}
\affiliation{Department of Physics, Virginia Tech, Blacksburg, VA 24061}


\date{\today}

\begin{abstract}
Adaptive variational quantum simulation algorithms use information from the quantum computer to dynamically create optimal trial wavefunctions for a given problem Hamiltonian. A key ingredient in these algorithms is a predefined operator pool from which trial wavefunctions are constructed. Finding suitable pools is critical for the efficiency of the algorithm as the problem size increases. Here, we present a technique called operator pool tiling that facilitates the construction of problem-tailored pools for arbitrarily large problem instances. By first performing an ADAPT-VQE calculation on a smaller instance of the problem using a large, but computationally inefficient operator pool, we extract the most relevant operators and use them to design more efficient pools for larger instances. We demonstrate the method here on strongly correlated quantum spin models in one and two dimensions, finding that ADAPT automatically finds a highly effective ansatz for these systems. %the resulting state preparation circuits \textcolor{blue}{are comparable or shorter compared to existing methods}. 
Given that many problems, such as those arising in condensed matter physics, have a naturally repeating lattice structure, we expect the pool tiling method to be a widely applicable technique apt for such systems.
\end{abstract}

%\keywords{}

\maketitle


%\subsection{}
%\subsubsection{}

Variational quantum eigensolvers (VQEs) comprise an important and much-studied class of algorithms for near-term quantum computers \cite{Peruzzo2014,Cerezo2021,Bharti2022,Tilly2022}. As examples of hybrid quantum-classical algorithms, they divide the computational task of optimizing an objective function $f(\vect{\theta})$ (for instance, the ground state energy of a physical system) between a quantum and a classical processor, in which the former efficiently queries the objective function, while the latter determines updated guesses for the variational parameters $\vect{\theta}$. In a VQE, the objective function takes the form of an expectation value $f(\vect{\theta})= \vev{H}_{\vect{\theta}}$ of an observable $H$ in the state $\ket{\psi(\vect{\theta})}$. The latter is prepared by a parameterized quantum circuit, and is subsequently measured to determine the relevant expectation value.

While the choice of $H$ is typically fixed by the problem at hand, the success of the algorithm depends strongly on the form of the circuit that prepares $\ket{\psi(\vect{\theta})}$ (also known as the ``ansatz'') and the optimization method used.  Many different ansatze have been proposed, including hardware-efficient \cite{Kandala2017} and physics- or chemistry-inspired \cite{Peruzzo2014,McClean2016,Wecker2015,Grimsley2019b} ones. In general, a good ansatz ought to be expressive enough to closely approximate the solution for suitable values of $\vect{\theta}$, while maintaining a low circuit depth, for compatibility with the limited coherence times of noisy quantum devices.

An important development in this regard was the introduction of the ADAPT-VQE method \cite{Grimsley2019a}, which put forward the idea of adaptive VQEs. This approach iteratively constructs an ansatz 
\begin{align}
|\psi_{n+1}(\vect{\theta}^{(n+1)})\rangle = e^{-i\theta_{n+1 }A_{n+1}}|\psi_{n}(\vect{\theta}^{(n)})\rangle,
\end{align}
where $\vect{\theta}^{(n)}=(\theta_1,\dots,\theta_n)$ are the variational parameters at step $n$ in the algorithm.  In adaptive VQEs, the operators $A_n$ are selected from a fixed, predefined ``operator pool'' according to a selection criterion. In the ADAPT-VQE approach, the operator chosen at each step is the one that maximizes the gradient of the objective function, i.e., the $A_k$ for which $\left| \left. \partial \langle H \rangle / \partial \theta_{n+1}\right|_{\theta_{n+1}=0} \right|$ is maximized, where $H$ is the objective function operator. Intuitively, this criterion seeks to identify the operators $A_k$ which are most effective at decreasing the objective function at a given step.

The ADAPT-VQE algorithm was first applied to finding the ground state energies of small molecules, where it obtained high accuracy with shallower circuit depths compared to other commonly used ansatze. The approach was subsequently generalized to a number of other applications, including optimization problems \cite{LZhu2020,Chen2022}, real and imaginary time evolution \cite{Yao2021,Gomes2021}, the preparation of excited states \cite{Zhang2021}, and strongly-correlated lattice models \cite{Gyawali2022}. The adaptive philosophy towards ansatz construction has also found other expressions, as in evolutionary algorithms and related methods \cite{Rattew2019,Chivilikhin2020}. 

As one might expect, the success of ADAPT is crucially dependent on the choice of operator pool. The first application of the method used pools consisting of fermionic operators that correspond to single and double excitations of electrons. It was then discovered that qubit-based pools, consisting of individual Pauli strings, can also reach the ground states of molecular systems but with considerably reduced counts of two-qubit gates \cite{Tang2021}. Subsequent work \cite{Yordanov2021} demonstrated that even more savings can be achieved by symmetrizing the pool of Ref.~\cite{Tang2021}. 

In general, the optimal choice of pool for a given problem is subject to a trade-off: on the one hand, pools with more operators are more likely to lead to convergence (and more quickly), due to the greater flexibility of operator selection at each step in the algorithm. On the other hand, since $|\partial \langle H \rangle / \partial \theta|$ is calculated for each operator in the pool at every step, this can lead to considerable measurement overhead when estimating the expectation values on a quantum computer. In seeking to reduce this, it was proven that ``minimally complete'' pools can be constructed that contain only $2L-2$ operators (where $L$ is the number of qubits) \cite{Tang2021}.  These pools have the property that all states in the Hilbert space can be obtained from an ansatz using only the operators of the given pool. However, due to the local nature of the ansatz updates, it is not guaranteed that the sequence of operators required can be found by the algorithm. One method for alleviating some of these issues is to explicitly encode information about symmetries of the problem into the choice of operator pool \cite{Shkolnikov2021}. More recently, it has been shown \cite{anastasiou2023really} that certain hardware-efficient pools containing up to ${\cal O}(L^4)$ operators only incur a ${\cal O}(L)$ measurement overhead on top of what is needed for standard, non-adaptive VQEs; this is the same measurement overhead scaling as for minimal complete pools, but without a reduction in operator selection flexibility.

The presence of symmetries in some systems leads to degeneracies in the gradient of the objective function amongst various operators in the pool. One strategy to address this involved adding multiple disjoint operators to the ansatz at each step, which was found to lead to significantly shallower circuits \cite{PAnastasiou2022tetrisadaptvqe}, and consequently a reduction in the overall measurement count as well.

Despite these advances, the space of possible choices for operator pools remains staggeringly large, growing exponentially in the number of qubits. At present, identifying ultimately successful pools is somewhat of an art, often based on a trial-and-error approach in which various candidates are investigated and compared, slowing the pace of discovery.

In this work, we present a general systematic procedure for constructing pools that are both small in size and also facilitate rapid convergence to the ground state. This is achieved by tiling successful pools from small problem instances to construct pools that are effective for larger problems. We then illustrate the method on the example of the spin-1/2 XXZ model, a prototypical correlated spin model. While we find that our approach leads to shorter or similar-depth state preparation ansatze compared to previous methods, our approach is a more general technique to systematically solve lattice problems rather than relying on ad hoc solutions. Finally, we discuss other variations of the underlying idea and their connection to many-body expansions used in quantum chemistry.

%As noted above, 
ADAPT instances that use larger pools are expected to converge more often and with fewer iterations, owing to the greater number of states that can be accessed at each step. Among qubit-based pools, the largest possible one is the ``full Pauli pool'', consisting of all Pauli strings on $L$ qubits, excluding the identity. As this pool includes all other qubit pools as proper subsets, it performs at least as well as each of them, in the sense of making the locally optimal choice.  However, the exponential scaling of the size, $4^L-1$, clearly prevents its application beyond small problem instances.

Many problems, such as those arising in condensed matter physics, have a simple repeated (lattice) structure that relates instances defined for different numbers of qubits, up to finite-size effects. Exploiting the lattice structure of these systems on near-term quantum devices has attracted increasing attention \cite{Liu2019,Eddins2022,Fujii2022}. Example problems include spin models (as used in the field of quantum magnetism), as well as fermionic ones like the Hubbard model, well-known from the study of high-temperature superconductivity. The lattice structure of these models suggests the following approach to designing effective pools: (1) Start with a small instance of the particular model or problem of interest on $L_1$ qubits and run ADAPT-VQE with a large, highly expressive pool that can quickly converge to the desired solution. To address the degenerate gradients that typically arise when using a sizeable operator pool, we run a large number of trials, growing the ansatz by picking randomly from the degenerate operators, thus sampling a variety of paths to convergence. (2) Take the set of operators chosen by ADAPT over all trials as the basis for a new pool defined on a larger instance of the same problem on $L_2>L_1$ qubits. This requires taking the tensor product of each operator with the single-qubit identity $L_2-L_1$ times, such that operators in the new pool have the correct dimension. Since the products can be applied either on the left or the right, a single operator of the original pool yields $L_2-L_1+1$ operators for the new one, corresponding to ``tiling'' the operator across the system in the one-dimensional case (with straightforward generalizations to higher dimensions). Importantly, while the original pool size can be exponential in the number of qubits $L_1$ (as with the full Pauli pool), the number of operators in the new pool scales linearly with the system size $L_2$. 

\begin{figure}
\includegraphics[width=\columnwidth]{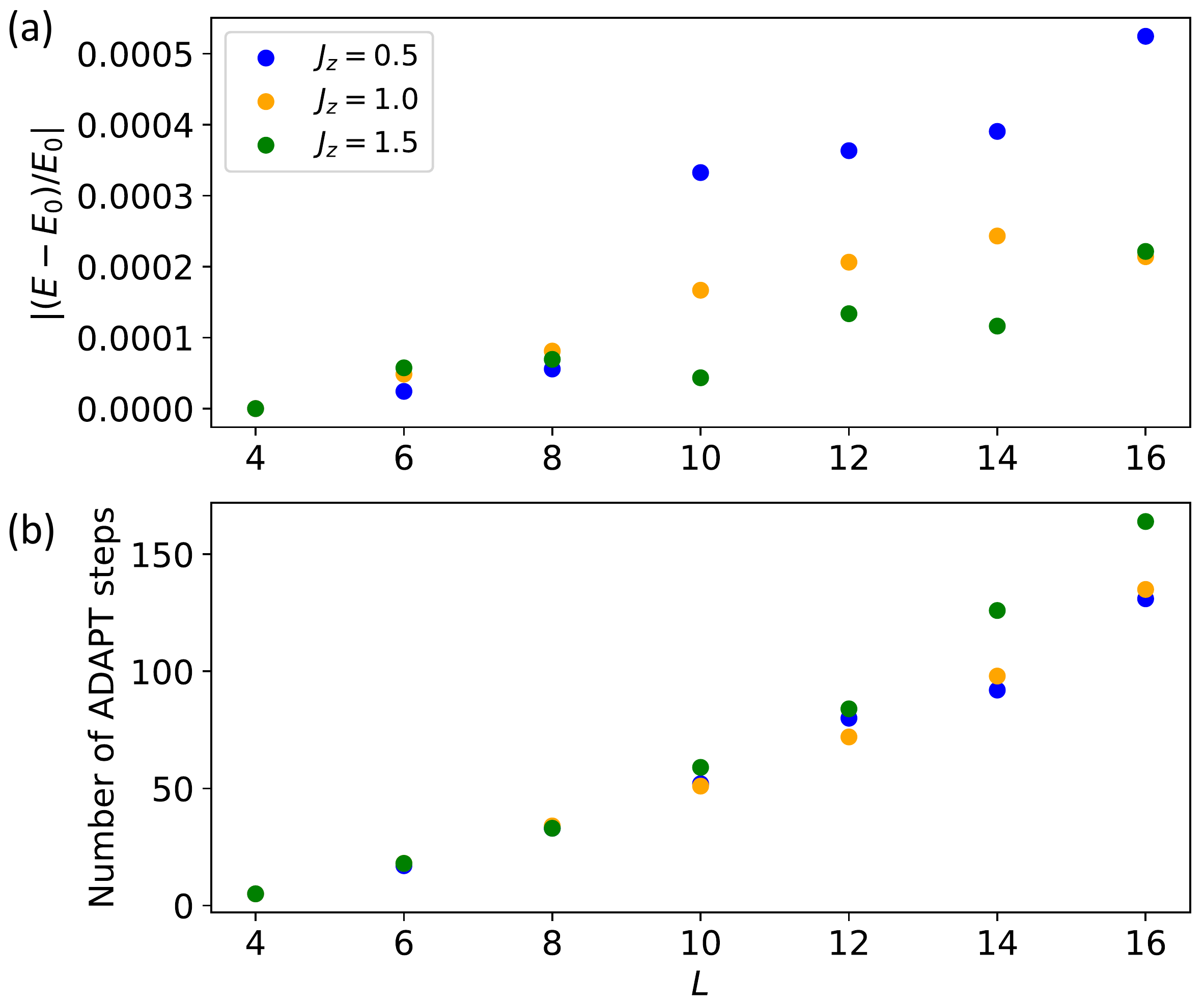}
\caption{(a) Relative error in the ground state energy versus system size $L$ for the XXZ chain. %using the Z-decorated XY pool. 
$E$ is the energy found by ADAPT after convergence, while $E_0$ is the exact ground state energy.  (b) Number of ADAPT steps required to converge the algorithm versus system size $L$. %, using the Z-decorated XY pool. 
The initial state is the N\'{e}el state $|\!\!\uparrow \downarrow \uparrow \downarrow \cdots \rangle$, and we set $\varepsilon=0.01$. Variational parameters are optimized using the LBFGS algorithm, with $\theta_{n+1}=0$ and $\{\theta_{j}\}_{j=1,\dots,n}$ set to the optimal values from the previous ADAPT iteration. \label{fig:enerconvergnumstepsvsL}}
\end{figure}

We now demonstrate the viability of this strategy through numerical simulations on a concrete example. We consider both  one-dimensional and two-dimensional spin-1/2 XXZ models. The first case consists of a chain of $L$ sites with open boundary conditions, whose Hamiltonian is given by
\begin{align}
H^{(1D)}_{XXZ} =  \sum_{i=1}^{L-1} J_{xy} \left( X_i X_{i+1} + Y_i Y_{i+1} \right) + J_z Z_i Z_{i+1},
\end{align}
where $X_i$, $Y_i$, and $Z_i$ are Pauli operators acting on site $i$. We set $J_{xy}=1$ throughout this work. For the first part of the tiling method, we solve the $L=3$ %$L=4$ 
case of the XXZ chain using the full Pauli pool. ADAPT always converges in three steps for this simple problem, and running trials to accommodate degenerate gradients as described above reveals that it chooses a set of operators that preserve time-reversal symmetry and simultaneously commute with the operator $Z_1 Z_2 Z_3$. The complete list of chosen operators is available in the Supplemental Material~\cite{SupplementalMaterial}. %: \texttt{IXY}, \texttt{IYX}, \texttt{XYI}, \texttt{XIY}, \texttt{YXI}, \texttt{YIX}, \texttt{ZXY}, \texttt{YZX}, \texttt{YXZ}, \texttt{XZY}, \texttt{ZYX}, \texttt{XYZ}} %\texttt{YXII}, \texttt{IIYX}, \texttt{ZYXI}, \texttt{XZYI}, and \texttt{YXZI} in sequential order (here the site indices have been suppressed and the identity is written explicitly). 
Convergence of the algorithm is defined to occur when the maximum gradient for a given step drops below a threshold $\varepsilon$ \cite{Grimsley2019a}. %We note that in obtaining this result, we used a modified operator selection compared to the original version of ADAPT: If multiple operators have the same maximal gradient at a given step, we select the one that has the lowest Pauli weight (i.e. the number of non-identity Pauli operators in the string). This is useful both for reducing the number of two-qubit gates (which is better suited for noisy devices) and for identifying particular motifs that can be tiled when going to a larger system. Indeed, given the simple structure of the set of operators chosen by ADAPT, we first generalize the starting $L=4$ pool to include other related Pauli strings, before tiling these to produce pools for larger systems. Thus we start from a ``Z-decorated XY pool'' consisting of the operators: \{\texttt{YX}, \texttt{XY}, \texttt{ZYX}, \texttt{ZXY}, \texttt{YZX}, \texttt{XZY}, \texttt{YXZ}, \texttt{XYZ}\}. 
We then tile these operators to create pools for systems with $L>3$ and run ADAPT. Fig.~\ref{fig:enerconvergnumstepsvsL}(a) shows the relative error in the ground state energy compared to the exact solution for this model, for system sizes $4 \leq L \leq 16$ and several values of the interaction strength $J_z$. In each case, the error is less than 0.06\%, indicating good convergence across a range of $J_z$ that includes both the paramagnetic and ordered regimes.

In Fig.~\ref{fig:enerconvergnumstepsvsL}(b) we present the number of ADAPT iterations required to find the ground state as a function of $L$. We note that this value is also equal to the number of optimization parameters in the final ansatz. The behavior over the simulated range appears to be close to polynomial. %linear, such that a simple extrapolation would suggest that $10^3$--$10^4$ parameters are needed for a system with $L \approx 100$ sites. 
To our knowledge, previous works that applied VQE to the XXZ model only focused on the case of closed boundary conditions. Ref.~\cite{BravoPrieto2020scalingof} used a fixed-circuit ansatz to study the $J_z = 0.5$ XXZ model for $L=12$ sites, finding that 168 parameters were required to obtain an energy accuracy of log$_{10}(1/|E-E_0|) \approx 1.0$, whereas in our calculations with the tiled pool, ADAPT achieves a higher accuracy of log$_{10}(1/|E-E_0|) \approx 1.9$, with an ansatz containing 132 variational parameters. On the other hand, for $J_z=1$ we find that ADAPT converges with 122 parameters, which is comparable to what Ref.~\cite{Wiersema2020} found using the Hamiltonian variational ansatz. These examples suggest that ADAPT with tiling can perform at least as well as previous approaches, if not better. Further evidence of this comes from Ref.~\cite{Jattana2022}, which examined various VQE ansatze for the XXZ model and found that the best-performing ansatz, which they termed the XY-ansatz, is constructed from operators with support on two or three qubits, similar to those in the tiled pool discovered by ADAPT.

\begin{figure}
\includegraphics[width=\columnwidth]{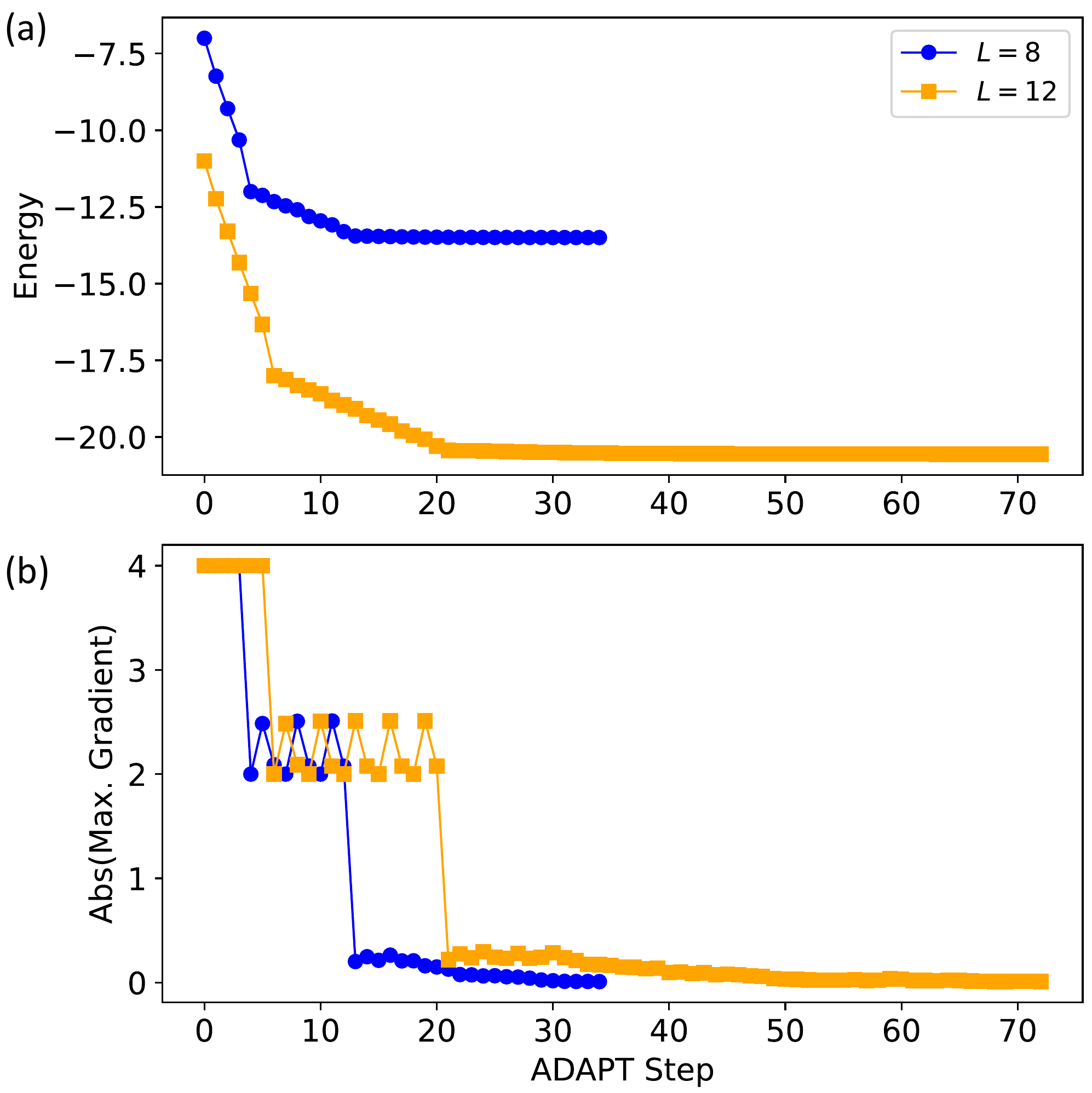}
\caption{(a) Energy and (b) absolute value of the maximum gradient as a function of ADAPT iteration number for the XXZ chain. %, using the Z-decorated XY pool. 
Simulation parameters are the same as for Fig.~\ref{fig:enerconvergnumstepsvsL} with $J_z=1$. \label{fig:energygradvsstep}} 
\end{figure}

We now turn to the details of how ADAPT-VQE approaches convergence in the present example. The energy and the absolute value of the maximum gradient at each iteration of the algorithm are shown in Fig.~\ref{fig:energygradvsstep}(a),(b) respectively. Three distinct regimes are apparent in both quantities. The energy decreases sharply during the first $L/2$ iterations, while the maximum gradient remains at its initial value for $L/2$ iterations. This is followed by a second region in which the energy decreases roughly linearly with iteration number, while the gradient shows a plateau with oscillations between several values. The final regime shows a slower decay both in the energy and the maximum gradient. One can obtain some physical insight into this behavior by inspecting the operators chosen by ADAPT and the corresponding optimal parameter values. The first $\frac{L}{2}$ operators in the ansatz are of the form $e^{-i \theta_k X_{2k-1} Y_{2k}}$ with $k=1,\dots,L/2$ and $\theta_k \approx - \pi/4$. Acting on the initial N\'{e}el state, this sequence produces a superposition of basis states in which pairs of up and down spins have been swapped, with a nontrivial relative sign structure between them. 
The other two regimes in the ADAPT convergence process involve the %Z-decorated 
three-local operators, and possess different optimal parameter values which do not form an easily discernible pattern. Nevertheless, it is clear that the operations performed during the first and second regimes are more effective for reducing the energy overall, while the final regime involves finer tuning of the wave function.

Importantly, the operators chosen by ADAPT for the \textit{small} problem instance span the symmetry subsector that respects time-reversal symmetry, conserves the expectation value of $Z_1 Z_2 \cdots Z_L$, and contains the reference state and the target state, implying that the tiled operators constructed from these operators span the relevant symmetry sector for the \textit{large} problem instance. This in turn demonstrates that the tiled pool is sufficiently expressible to represent the target state. A more detailed proof is available in the Supplemental Material~\cite{SupplementalMaterial}.

\begin{figure}
\includegraphics[scale=0.35]{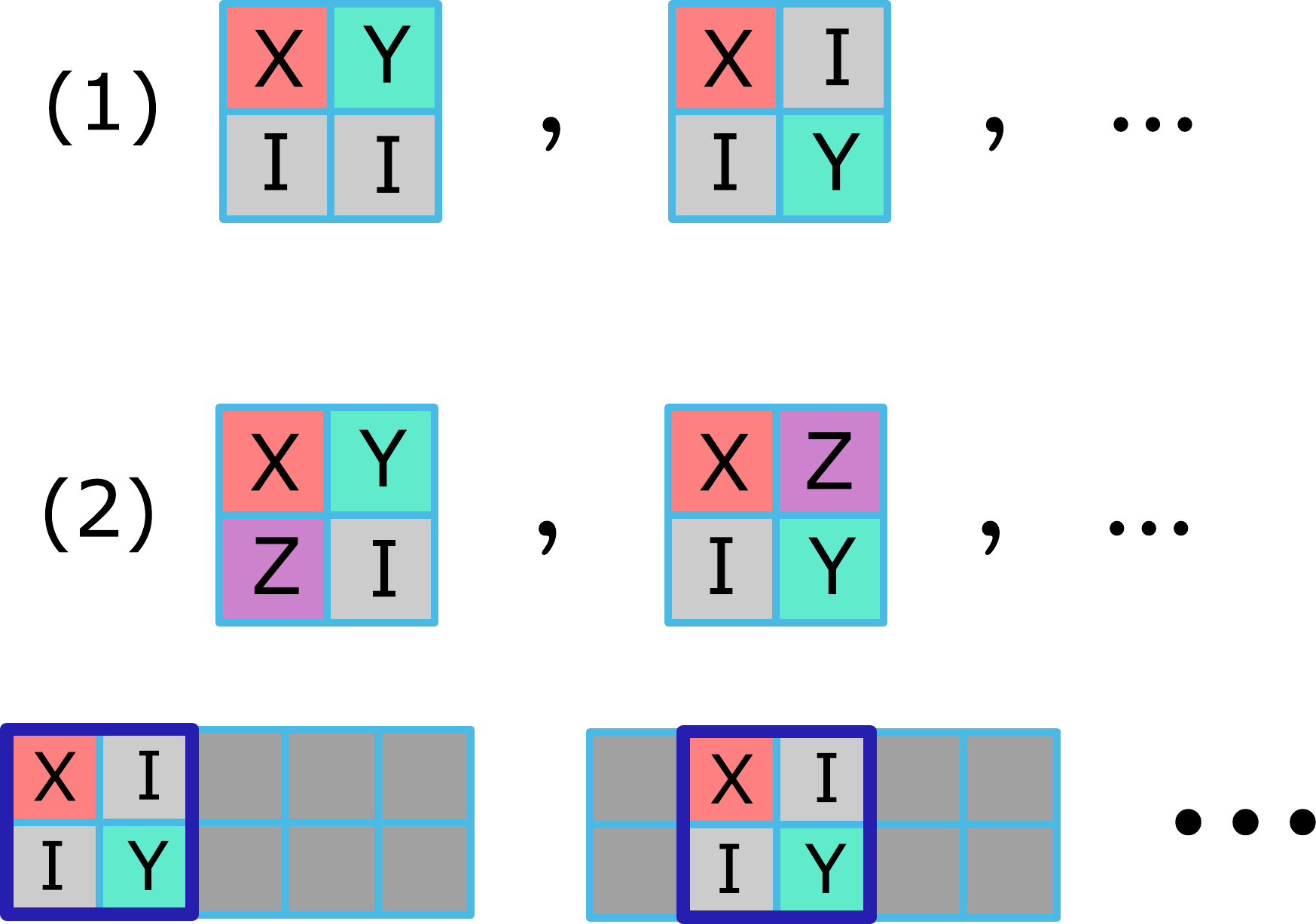}
\caption{Illustration of the tiling procedure for the two-dimensional XXZ model. 
\label{fig:2dpool}} 
\end{figure}

For the two-dimensional XXZ model we study the case of open boundary conditions and nearest-neighbor interactions on rectangular lattices of $N = L_x L_y$ sites:
\begin{equation}
H^{(2D)}_{XXZ} =  \sum_{\langle i,j \rangle} J_{xy} \left( X_i X_{j} + Y_i Y_{j} \right) + J_z Z_i Z_{j},
\end{equation}
where $\langle i,j \rangle$ denotes nearest neighbors. In generating the ADAPT operator pool, it is natural to consider tiles consisting of 2$\times$2 blocks, given the two-dimensional nature of the problem. Inspired by the successful strategy of collecting operators from a number of ADAPT runs on the tile in the one-dimensional case, we run several trials, randomly picking operators in the case of degenerate gradients. In this case, ADAPT always converges in either five or six steps, choosing a set of operators that simultaneously satisfy time-reversal symmetry and commute with $Z_1 Z_2 Z_3 Z_4$, and importantly, form a complete basis for the symmetry subsector conserving the two. %from the following operators: \texttt{IIXY}, \texttt{IIYX}, \texttt{IXYZ}, \texttt{IXZY}, \texttt{IYXZ}, \texttt{IYZX}, \texttt{IZXY}, \texttt{IZYX}, \texttt{XIYZ}, \texttt{XIZY}, \texttt{XYII}, \texttt{XYIZ}, \texttt{XYZI}, \texttt{XYZZ}, \texttt{XZIY}, \texttt{XZYI}, \texttt{YIXZ}, \texttt{YIZX}, \texttt{YXII}, \texttt{YXIZ}, \texttt{YXZI}, \texttt{YXZZ}, \texttt{YZIX}, \texttt{YZXI}, \texttt{ZIXY}, \texttt{ZIYX}, \texttt{ZXIY}, \texttt{ZXYI}, \texttt{ZYIX}, \texttt{ZYXI}, \texttt{ZZXY}, \texttt{ZZYX}.} 
These blocks are then tiled across the ladder (in overlapping fashion) to create the operator pool. The procedure is illustrated in Fig.~\ref{fig:2dpool}

\begin{figure}
\includegraphics[width=\columnwidth]{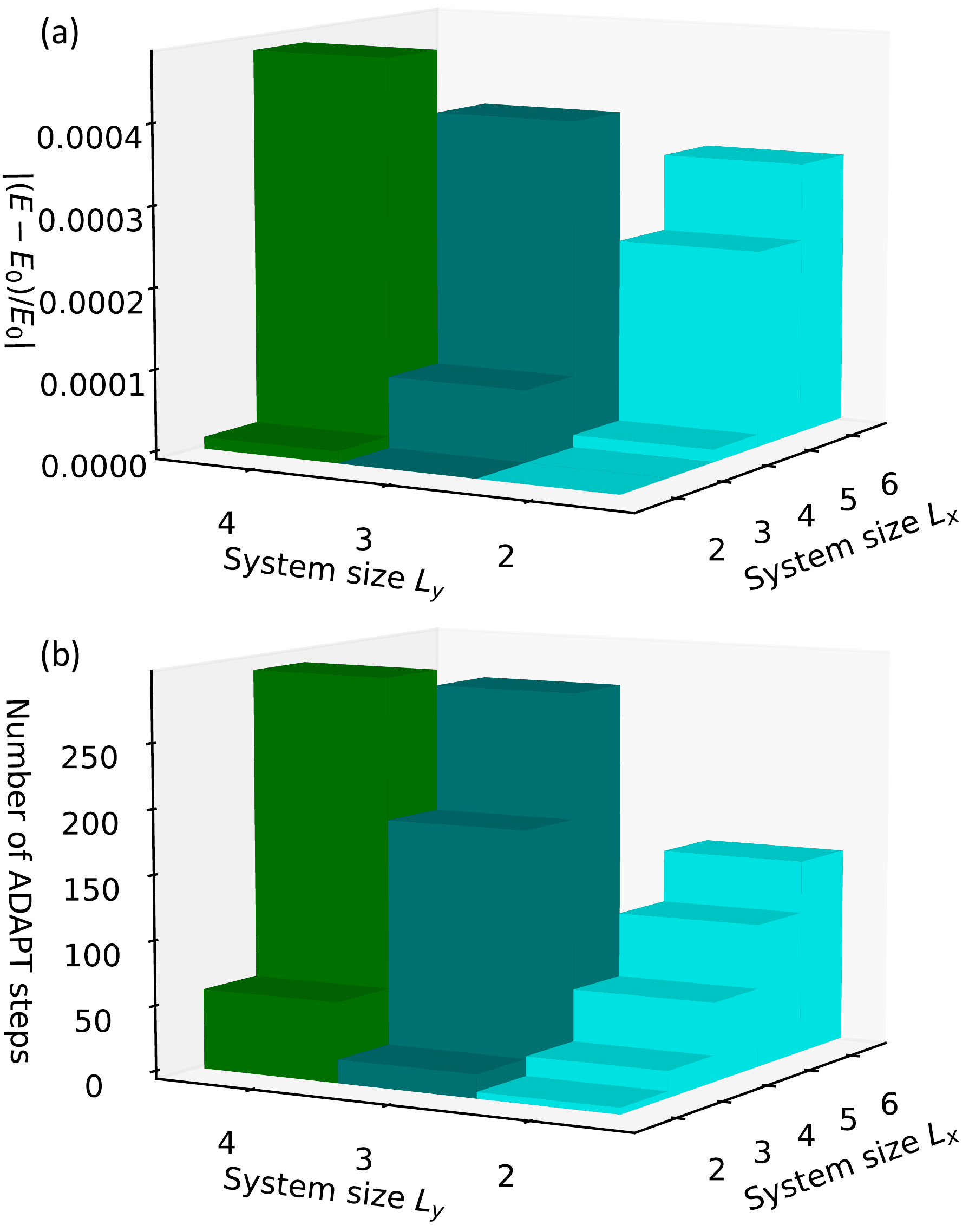}
\caption{(a) Relative error in the ground state energy and (b) number of ADAPT steps required for convergence for the 2D XXZ model with a square geometry. %, using the pool described in Fig. \ref{fig:2dpool}. 
The simulation parameter is $J_z=1$. \label{fig:enerconvergnumstepsvsLxLy}} 
\end{figure}

In Fig.~\ref{fig:enerconvergnumstepsvsLxLy} we present the energy error [Fig.~\ref{fig:enerconvergnumstepsvsLxLy}(a)] and number of steps to convergence [Fig.~\ref{fig:enerconvergnumstepsvsLxLy}(b)] for %the 2D XXZ model
this case. While the ground state is accurately approximated here as well, the computational difficulty grows rapidly with increasing $L_x$ and $L_y$. %associated with the $L_x^2$ growth in the total number of sites limits our calculations to only three values ($L_x=2,3,4$). 
Although a quantitative extrapolation to large systems would not be warranted, it is clear that the required number of ADAPT steps grows more quickly with system size than in the 1D case. Notably, this nonlinear growth remains when the number of steps is plotted as a function of $L_x\times L_y$, suggesting that the computational difficulty is at least partially due to the intrinsic connectivity of the 2D lattice, and not just the total number of sites. In any case, it appears that the 2D XXZ model presents a significant challenge for VQE methods in the quantum advantage regime, due to the large number of optimization steps needed. Nevertheless, the rapid progress in variational quantum algorithms lends hope that such difficulties may be surmounted.

Having demonstrated the power of the ``tiling'' approach to designing operator pools for ADAPT, we turn to possible generalizations of the method. While the above examples focused on 1D and 2D systems, it is clear how to generalize to higher dimensions (though likely to be computationally challenging). Generalizations to other popular models in condensed matter physics, such as the Hubbard model, are also straightforward. While such constructions are quite natural for lattice systems with short-range interactions, it is less clear that a simple tiling procedure will work for more complex geometries and/or long-range interactions (as found in quantum chemistry applications, for instance). In this case, one can consider the more general approach of starting from the full problem, selecting out various subsets of the degrees of freedom, and solving these reduced problems using ADAPT with a large operator pool. The resulting operators chosen from each subproblem can then be collected to generate a pool for executing ADAPT on the original problem. This method bears a close resemblance to many-body expansion techniques used in quantum chemistry~\cite{Gros1993,Senechal2000}. We leave further development of these ideas for future work.

To conclude, we have proposed and demonstrated a systematic operator tiling method for designing small but accurate operator pools for use with the ADAPT-VQE algorithm. The strength of this approach lies in its versatility in producing efficient representations of the ground state for lattice models in general, and is a significant advantage for using ADAPT on noisy quantum devices. We have illustrated the approach on the open XXZ spin chain and notably, we found that the tiling method automatically found a highly effective ansatz for this system~\cite{Jattana2022}. %favorably compared to the popular Hamiltonian variational ansatz, requiring a significantly reduced number of variational parameters to reach convergence. 
Confirming the usefulness of the operator tiling method for other models in strongly correlated physics is an interesting direction for future work.

\begin{acknowledgments}
Numerical calculations of the exact ground state energies were performed with the QuSpin Python library \cite{Weinberg2017}. This work was supported by the Department of Energy. E.B. and N.J.M. acknowledge Award No. DE-SC0019199, and S.E.E. acknowledges the DOE Office of Science, National Quantum Information Science Research Centers, Co-design Center for Quantum Advantage (C2QA), contract number DE-SC0012704.
\end{acknowledgments}

\section{Supplemental Material for ``Scaling adaptive quantum simulation algorithms via operator pool tiling''}

\subsection{Outline of proof of completeness of the tiled pool}
Here we outline how to prove that for a translation-symmetric Hamiltonian on a lattice, given reference states in the appropriate symmetry sectors, if we start from a complete operator pool (i.e., the operators span the entire Lie algebra in the relevant symmetry sector) on a small problem instance (the `tile') on $L_1$ qubits, then the tiled pool for a larger problem instance on $L_2>L_1$ qubits, generated using the operators chosen by ADAPT-VQE~\cite{Grimsley2019a} for the tile, will also be complete. Completeness of the pool in turn implies that the pool has sufficient expressibility to represent any state in the relevant symmetry sector of the $L_2$-qubit system~\cite{Tang2021,Shkolnikov2021}, including the target state. 

We begin by summarizing our procedure for constructing the pool on $L_1$ qubits as described in the main text. First, consider a Hamiltonian $H$ for a small problem instance on $L_1$ qubits along with a highly expressive operator pool of Pauli string-type operators and a given reference state. At each step, ADAPT adds to the ansatz the operator with the highest gradient among the pool operators, growing the ansatz until convergence is achieved. To account for the possibility of multiple operators having identical gradients, several ADAPT-VQE calculations are run in which the algorithm randomly picks operators from those with degenerate gradients to add to the ansatz at each step. Thus, various routes to convergence are sampled. The chosen operators obtained from these runs are collected and stored. Note that if the system has a conserved quantity $S$ that commutes with the Hamiltonian, and if we start from a reference state $|\psi_{init}\rangle$ which lies in the same symmetry subsector as the target state, then ADAPT will only pick operators that respect this symmetry~\cite{Shkolnikov2021}. 

For all the examples considered in the main text, we find numerically that ADAPT chooses a set of operators that generates the \textit{entire} Lie algebra of the symmetry subsector, i.e., these operators constitute a complete pool for the small problem on $L_1$ qubits. For other lattice problems, one can of course easily check whether the set of operators collected in this way forms a complete pool on $L_1$ qubits. Assuming this is the case, we will now outline how to prove that the set of tiled pool operators on $L_2$ qubits, which are obtained by padding the original set with $L_2-L_1$ identity operators in all possible ways without breaking the tile, necessarily form a complete pool on $L_2$ qubits.

To start, consider the concrete example of a linear qubit array with $L_1=3$ and $L_2=4$. There are then two types of tiled pool operators that we construct for the 4-qubit system by adding identities to the three-qubit pool obtained from ADAPT: 
$I\otimes\sigma_a\otimes\sigma_b\otimes\sigma_c$ and $\sigma_d\otimes\sigma_e\otimes\sigma_f\otimes I$, where $a,b,c,d,e,f \in \{0,1,2,3\}$, with $\sigma_0=I$, $\sigma_1=X$, $\sigma_2=Y$, $\sigma_3=Z$. Here, we assume that the sets of operators $\sigma_a\otimes\sigma_b\otimes\sigma_c$ and $\sigma_d\otimes\sigma_e\otimes\sigma_f$ generate the full Lie subalgebra of Pauli strings on 3 qubits that commute with $S$. If the Hamiltonian also possesses time-reversal symmetry (i.e., the Hamiltonian is real),
then the subalgebra would be further restricted to contain only odd Pauli strings (operators that contain an odd number of $Y$'s) that commute with $S$, since in this case the eigenstate wavefunctions can always be chosen real. The question is then whether we can generate any 4-qubit (odd) Pauli string that commutes with $S$ from commutators of tiled operators like $I\otimes\sigma_a\otimes\sigma_b\otimes\sigma_c$ and $\sigma_d\otimes\sigma_e\otimes\sigma_f\otimes I$. (We assume that $S$ can be readily extended to the larger 4-qubit system in accordance with the translation symmetry of the Hamiltonian.) In order to prove that this is indeed the case, we need to first specify a particular Hamiltonian $H$ and symmetry $S$. We prove this for the XXZ models considered in this work in the next section below. We then use this completeness result for two overlapping tiles (which is the $L_2=4$ case in our linear $L_1=3$ example) as the base case for an inductive proof that extends the same result to any $L_2$. 

Let us illustrate how the inductive proof works by continuing on to the case of $L_2=5$ for our linear $L_1=3$ example.
Assuming that it is possible to generate any symmetry-preserving 4-qubit Pauli string from the $L_1=3$ tiled pool, the next question is whether we can generate any 5-qubit symmetry-preserving Pauli string in a similar fashion, where now $L_2=5$. We could try to form 5-qubit strings by taking commutators of operators like $I\otimes I\otimes\sigma_a\otimes\sigma_b\otimes\sigma_c$ and $\sigma_d\otimes\sigma_e\otimes\sigma_f\otimes\sigma_g\otimes I$. If the $L_2=4$ case holds, then it immediately follows that 5-qubit strings for which the 4-qubit substring on qubits 2 through 5 is symmetry-preserving can be generated via such commutators. However, if the 4-qubit substring is not symmetry-preserving, then further analysis is needed. In particular, we would need to show that any symmetry-violating 4-qubit Pauli string can be produced from nested commutators of the form
\begin{equation}\label{eq:nestedcommutators}
[...[[V,P_1],P_2]...],
\end{equation}
where $V=\sigma_e\otimes\sigma_f\otimes\sigma_g\otimes I$ violates either time-reversal symmetry or $S$ symmetry or both, while each of the $P_i=\sigma_{a_i}\otimes\sigma_{b_i}\otimes\sigma_{c_i}\otimes\sigma_{d_i}$ preserves both symmetries, where at least one of $\sigma_{a_i}$ and $\sigma_{d_i}$ is the identity operator. Note that the nested commutator above has the same parity as $V$ and it commutes with $S$ if and only if $V$ commutes with $S$. Thus, showing that any symmetry-preserving 5-qubit Pauli string can be generated from commutators of tiled pool operators is tantamount to showing that, given a $V$ of a certain parity that commutes or anticommutes with $S$, the above nested commutators span the entire set of operators that share the same parity and commutativity relation with $S$. 

The above argument can in fact be extended to symmetry-preserving Pauli strings on any number of qubits $L_2$ via induction. If we can generate any symmetry-preserving string on $L_2$ qubits, then we can obtain any symmetry-preserving string on $L_2+1$ qubits by taking nested commutators of an $L_2$-qubit string with tiled pool operators that have support only on the last 4 qubits, $L_2-2,...,L_2+1$. Thus, in order to prove that the tiled pool is complete, it suffices to show that any 4-qubit Pauli string can be generated from nested commutators as in Eq.~\eqref{eq:nestedcommutators} for some choice of $V$ and the $P_i$.   

Although here we focused on a linear array of qubits with $L_1=3$, this argument works regardless of the dimension or connectivity of the lattice, and regardless of the size $L_1$ of the tiles, so long as the requisite conditions on overlapping tiles are satisfied. More generally, one would need to show that any Pauli string with support on the union of two overlapping tiles can be generated by taking nested commutators of some Pauli string $V$ that has support on one tile with tiled pool operators $P_i$ that each have support on one of the two tiles.
Since this proof does not require assumptions about the form of the Hamiltonian, we expect it to be applicable to a wide variety of models. In the following section, we show explicitly that the above conditions hold for the tiled pool for the XXZ model, and that this pool is thus complete for any number of qubits $L_2$. 

\subsection{Proof of tiled pool completeness for the XXZ model}
For the spin models considered in the manuscript, the operator 
$P_Z = Z_1 Z_2 \cdots Z_n$
(where $n$ is the length of the spin system) commutes with the Hamiltonian. In addition to $P_Z$, the Hamiltonian has time-reversal symmetry $\mathcal{T}$.

Upon running ADAPT calculations on this system using the Néel antiferromagnetic state as a reference state and the full Pauli pool as the operator pool, we find that ADAPT chooses a set of operators that generate the Lie algebra for the subspace that simultaneously preserves the expectation value of $P_Z$ and respects $\mathcal{T}$. For a one-dimensional tile of size $L_1 = 3$ with the Néel antiferromagnetic state as the reference state, ADAPT chooses the operators \{$IXY$, $IYX$, $XYI$, $YXI$, $ZXY$, $YXZ$, $ZYX$, $XYZ$\} for $0.0 \leq J_z < 1.0$ and \{$IXY$, $IYX$, $XYI$, $XIY$, $YXI$, $YIX$, $ZXY$, $YZX$, $YXZ$, $XZY$, $ZYX$, $XYZ$\} for $J_z \geq 1.0$. Note that both these sets are complete in the relevant symmetry sector. For two-dimensional systems, we use a tile of size 2$\times $2, and find that for $J_z=1.0$, ADAPT chooses the set of operators: $IIXY$, $IIYX$, $IXYZ$, $IXZY$, $IYXZ$, $IYZX$, $IZXY$, $IZYX$, $XIYZ$, $XIZY$, $XYII$, $XYIZ$, $XYZI$, $XYZZ$, $XZIY$, $XZYI$, $YIXZ$, $YIZX$, $YXII$, $YXIZ$, $YXZI$, $YXZZ$, $YZIX$, $YZXI$, $ZIXY$, $ZIYX$, $ZXIY$, $ZXYI$, $ZYIX$, $ZYXI$, $ZZXY$, $ZZYX$. These operations, in one and two dimensions respectively, generate through their Lie algebras all Pauli string operators for the respective system sizes that preserve $P_Z$ and $\mathcal{T}$.

We next explicitly demonstrate the ability of the tiled operator pool to generate symmetry-preserving Pauli strings of any size via commutators of the tiled operators, focusing on the one-dimensional system. We do this by following the general strategy outlined in the previous section. That is, we first need to show that we can produce any symmetry-preserving 4-qubit Pauli string from commutators of tiled operators. We then need to show that we can transform a given symmetry-violating 4-qubit substring $V$ into any other symmetry-violating substring via such commutators as in Eq.~\eqref{eq:nestedcommutators}, as this ensures that we can generate symmetric Pauli strings on more than 4 qubits. For example, we could use $[XZYII,IIXYI]\propto XZZYI$ as our initial Pauli string on $L_2=5$ qubits, in which case the 4-qubit substring $V=ZZYI$ violates $P_Z$ symmetry but not time-reversal. We then need to show that this $V$ can be transformed into any other 4-qubit Pauli string that violates $P_z$ but preserves time-reversal using commutators with tiled operators (Eq.~\eqref{eq:nestedcommutators}) so that we obtain all symmetry-preserving 5-qubit Pauli strings that start with $X$ on the first qubit.

%The tiled operators are constructed by multiplying the operator tiles with single-qubit identity operators. 
Consider the commutator of a pair of 4-qubit tiled operators constructed from 3-qubit operator tiles with overlapping support,
\begin{equation}\label{four-qubit-comm}
    [ I \otimes \sigma_a \otimes \sigma_b \otimes \sigma_c,  \sigma_d \otimes \sigma_e \otimes \sigma_f \otimes I].
\end{equation}
Here, $a,b,c,d,e,f \in \{0, 1,2,3\}$ such that $\sigma_a \otimes\sigma_b\otimes \sigma_c$ and $\sigma_d \otimes\sigma_e\otimes \sigma_f$ are odd Pauli strings (containing one or three $Y$'s) and commuting with $Z\otimes Z\otimes Z$. Since the only possible three-qubit Pauli string containing three $Y$'s fails to commute with $Z\otimes Z\otimes Z$, the strings necessarily contain a single $Y$. Thus, the indices $\{a,b,c\}$ and $\{d,e,f\}$ are permutations of either $(0,1,2)$ or $(1,2,3)$. Next, if the two operators fail to commute, their commutator is proportional to their product, $\sigma_d \otimes (\sigma_a \sigma_e) \otimes (\sigma_b\sigma_f) \otimes \sigma_c$. Furthermore, their product commutes with $Z\otimes Z\otimes Z\otimes Z$, since they individually commute with it. Letting $\{a,b,c\}$ and $\{d,e,f\}$ in Eq.~\eqref{four-qubit-comm} take on all possible combinations of the values discussed above, as well as each of the operators in the commutator themselves, and including the higher-order commutators, it is straightforward to show that one obtains exactly the set of 4-qubit Pauli strings that commute with $Z\otimes Z\otimes Z\otimes Z$ and contain an odd number of $Y$'s. Thus, we have shown that the tiled operator pool can represent any state in the relevant symmetry sector for a system of $L_2=4$ qubits.

Next, consider the commutator of a pair of 5-qubit tiled operators---one constructed from a 3-qubit operator tile, and the other from a 4-qubit operator obtained in the previous step. This can be represented as
\begin{equation}\label{five-qubit-comm}
    [ I \otimes I \otimes \sigma_a \otimes \sigma_b \otimes \sigma_c , \sigma_d \otimes \sigma_e \otimes \sigma_f \otimes \sigma_g \otimes I].
\end{equation}
If the commutator is nonzero, then it is proportional to the product of the two operators. Now, $\{a,b,c\}$ must be a permutation of either $(0,1,2)$ or $(1,2,3)$ as in the earlier step. We will consider three separate scenarios for $\sigma_d\otimes \sigma_e\otimes \sigma_f\otimes \sigma_g\otimes I $, ensuring that it is overall odd and commutes with $Z\otimes Z\otimes Z\otimes Z\otimes Z$: (i) First, let $d = 0 \text{ or } 3$. In this case, $\sigma_e \otimes\sigma_f \otimes\sigma_g $ must be an odd Pauli string commuting with $Z\otimes Z\otimes Z$---i.e., it reduces to the 4-qubit case, and $\{e,f,g\}$ can take on the same values as $\{a,b,c\}$. (ii) Next, suppose $d=1$; now, $\sigma_e \otimes\sigma_f\otimes \sigma_g $ must be an odd Pauli string \textit{anti}-commuting with $Z\otimes Z\otimes Z$. This restricts $\{e,f,g\}$ to permutations of $(2,0,0)$, $(2,1,1)$, $(2,3,3)$, $(2,0,3)$ and $(2,2,2)$. (iii) Last, let $d=2$: here, $\sigma_e \otimes\sigma_f\otimes \sigma_g $ must be an even Pauli string \textit{anti}-commuting with $Z\otimes Z\otimes Z$. $\{e,f,g\}$ must then take on permutations of the values $(0,0,1)$, $(1,3,3)$, $(0,1,3)$, $(1,1,1)$, $(1,2,2)$. Letting $\{a,b,c\}$ and $\{d,e,f,g\}$ take on the appropriate values as detailed in each scenario, and including higher-order commutators, it is straightforward to show that one obtains all 5-qubit operators conserving $P_Z$ and $\mathcal{T}$.

Finally, consider the commutator of a pair of 6-qubit tiled operators,
\begin{equation}\label{six-qubit-comm}
    [ I \otimes I \otimes I \otimes \sigma_a \otimes \sigma_b \otimes \sigma_c , \sigma_d \otimes \sigma_e \otimes \sigma_f \otimes \sigma_g \otimes \sigma_h \otimes I].
\end{equation}
We previously showed that we obtain all 5-qubit operators in the relevant symmetry sector for $\sigma_e \otimes\sigma_f \otimes\sigma_g$ in Eq.~\eqref{five-qubit-comm} (i) odd and commuting with $Z\otimes Z\otimes Z$, (ii) odd and anti-commuting with $Z\otimes Z\otimes Z$, and (iii) even and anti-commuting with $Z\otimes Z\otimes Z$. In Eq.~\eqref{six-qubit-comm}, we can use the same argument to show that for all choices of $\{d,e\}$ such that $\sigma_f \otimes\sigma_g\otimes \sigma_h$ falls into one of the three cases (i)-(iii), the result will be a 6-qubit operator in the appropriate symmetry sector. This leaves us with one case---that of $\sigma_f \otimes\sigma_g \otimes\sigma_h$ in Eq.~\eqref{six-qubit-comm} being even and commuting with $Z\otimes Z\otimes Z$, which restricts $\{f,g,h\}$ to the values $(1,1,0)$, $(1,1,3)$, $(0,0,0)$, $(0,0,3)$, $(0,3,3)$, $(3,3,3)$, $(0,2,2)$ and $(2,2,3)$. Allowing $\{a,b,c\}$ and $\{d,e,f,g,h\}$ to take on the corresponding values, and including higher-order commutators, it is again straightforward to show that one obtains all 6-qubit operators conserving $P_Z$ and $\mathcal{T}$.

By induction, it follows that symmetry-preserving Pauli strings of any size can therefore be obtained from commutators of the tiled operators. The resultant operator pool spans the Lie algebra for the Hilbert space sector in which $P_Z$ and $\mathcal{T}$ are preserved, and in which the target state lies. Thus, we have shown how to construct an efficient operator pool for ADAPT-VQE calculations on translation-symmetric systems that is sufficiently expressive to reach the target state.


% uncomment for PRX Quantum formatting
\bibliography{supp}

\end{document}